\documentstyle[12pt]{article}
\parskip=\medskipamount

\newcommand {\cD}{{\cal D}}

\newcommand {\cF}{{\cal F}}

\newcommand {\cH}{{\cal H}}

\def\a{\alpha}
\def\b{\beta}

\def\d{\delta}
\def\e{\epsilon}

\def\G{\Gamma}

\def\l{\lambda}

\def\o{\omega}

\def\q{\theta}

\def\F{\Phi}

\def\L{\Lambda}

\newcommand{\ad}{{\dot{\alpha}}}                           %new
\newcommand{\ve}{\varepsilon}                            %new
\newcommand{\pa}{\partial}                           %new
\newcommand{\hf}{\frac12}

\newcommand{\be}{\begin{equation}}
\newcommand{\ee}{\end{equation}}
\newcommand{\bea}{\begin{eqnarray}}
\newcommand{\eea}{\end{eqnarray}}
\newcommand{\non}{\nonumber}
\begin{document}
%%%%%%%%%%%%%%%%%%%%%%%%%
%%%%%%%%%%%%%%%%%%%%%%%%

\begin{titlepage}
\thispagestyle{empty}

\begin{flushright}
TP-TSU-15/98 \\
hep-th/9810239 \\
October, 1998
\end{flushright}

\vspace{1cm}
\begin{center}
{\large\bf 
Non-holomorphic effective potential in N = 4 SU(n)  SYM}
\end{center}

\begin{center}
{\bf E. I.  Buchbinder}${}^\dag$, {\bf I. L. Buchbinder}${}^\ddag$ and {\bf  S. M. Kuzenko}${}^\P$
\footnote{On leave from: Department of Physics, Tomsk State University,
Tomsk 634050, Russia.\\ Supported in part by Deutsche Forschungsgemeinschaft.}\\
${}^\dag$ \footnotesize{
{\it  Department of Physics,
Tomsk State University\\
Tomsk 634050, Russia}}\\
\vspace{3mm}

${}^\ddag$ \footnotesize{
{\it Department of Theoretical Physics, 
Tomsk State Pedagogical University\\
Tomsk 634041, Russia}}\\
\vspace{3mm}

${}^\P$ \footnotesize{
{\it Sektion Physik, Universit\"at M\"unchen\\
Theresienstr. 37, D-80333 M\"unchen, Germany}}       \\

\end{center}
\vspace{1cm}

\begin{abstract}
We compute the one-loop non-holomorphic effective potential  
for  the $N=4$ $SU(n)$
supersymmetric Yang-Mills theory with the gauge symmetry 
broken down to the maximal torus $U(1)^{n-1}$. 
Our approach remains powerful for arbitrary gauge groups and
is based on the use of $N=2$ harmonic superspace formulation 
for general $N=2$ Yang-Mills theories  along with the superfield 
background field method. 
\end{abstract}
\vfill

\end{titlepage}

\newpage
\setcounter{page}{1}
\newpage
${}$\\
\noindent
Extended supersymmetry imposes  strong restrictions on the structure  
of  quantum field theories. One of the most prominent examples where
extended supersymmetry has played a substantial role is the exact
solution for non-perturbative low-energy effective action in the $N=2$ 
$SU(2)$ supersymmetric Yang-Mills theory given by Seiberg and Witten \cite{1}.
Their construction was generalized to arbitrary gauge
groups in  \cite{2}.  Another intriguing example  
comes from the $N=4$ Yang-Mills theory where 
the powerful symmetry properties allow one to 
exactly compute some Green's  
functions (see  \cite{hsw} and references therein).

In the background field formulation, the effective action of
$N=2$, $D=4$ super Yang-Mills theories is a manifestly gauge
invariant and  supersymmetric functional of the covariantly chiral
strength $W$ and its conjugate $\bar{W}$ \cite{3}.
In the Coulomb branch the effective action is in general reads
\begin{equation}
\Gamma[W,\bar{W}] ~= ~{\rm Im} \int {\rm d}^4x {\rm d}^4\theta\, {\cal F}(W) ~+~
\int {\rm d}^4x {\rm d} ^8\theta\,  {\cal H}(\bar{W},W)~+ ~\dots
\label{1}
\end{equation}
where the first term is integrated over the chiral subspace  of $N=2$ superspace
while the second term is integrated  over the full $N=2$ superspace parametrized by
$z^M \equiv(x^m, \theta_i^\alpha,\bar{\theta}^i_{\dot{\alpha}})$.
The elipsis denotes all terms involving  derivatives of the
strengths. The holomorphic potential ${\cal F}(W)$ 
dominates at low energies and presents itself the main object
of the Seiberg-Witten theory.
The non-holomorphic potential ${\cal H}(\bar{W},W)$ 
constitutes the next-to-leading finite quantum correction.
For finite $N=2$ Yang-Mills theories with matter,
$\cF (W)$ coincides with the classical gauge action,
and hence $\cH (\bar W, W)$ is the dominant quantum correction.
An important representative of such superconformal models is 
the $N=4$ Yang-Mills theory 
(the first finite quantum field theory found \cite{finite})
in which the $N=2$ matter is realized 
by a single hypermultiplet in the adjoint representation.

In a recent paper \cite{4} Dine and Seiberg 
showed that the requirement of scale and chiral invariance severely 
restricts the possible structure of ${\cal H}(\bar{W},W)$ in the $N=4$ Yang-Mills theory.
For the  $N=4$ $SU(2)$ theory broken down to $U(1)$, they found out the only
admissible form for the one-loop non-holomorphic potential:
\begin{equation}
{\cal H}(\bar{W},W)~=~c\;{\rm ln}\frac{\bar{W}^2}{\Lambda^2}\,
{\rm ln}\frac{W^2}{\Lambda^2}
\label{2}
\end{equation}
with some numerical coefficient $c$ and some scale $\Lambda$. 
The corresponding action given by the second term in (\ref{1}) 
turns out to be independent on $\L$.
Moreover, Dine and Seiberg argued  that ${\cal H}(\bar{W},W)$  gets
neither higher-loop perturbative nor instanton corrections,
what was confirmed by instantons calculations \cite{instanton} and two-loop
supergraph analysis \cite{11}. The problem
of explicit calculation of  the coefficient $c$ has been recently solved in Refs.
\cite{pvu,grr,6} on the base of different techniques, the final result being
$c=(8\pi)^{-2}$.  This value for $c$ was given  in \cite{pvu} to be 
the result of calculations based on the use of $N=1$ superspace 
formulation for the $N=4$ Yang-Mills theory. 
Gonzalez-Rey and Ro\v{c}ek \cite{grr} computed,
in the framework  of $N=2$ projective superspace approach, 
a special sector of the hypermultiplet low-energy action and 
then gave some grounds that the non-holomorphic 
effective potential  ${\cal H}(\bar{W},W)$
should have the same functional form.
Finally, in our paper \cite{6} we directly analysed, in the 
framework of $N=2$ 
harmonic superspace approach, the effective action
corresponding to the $N=2$ gauge multiplet of the full $N=4$  
Yang-Mills theory.

In the present paper we extend the
results of our work \cite{6}  to the case of 
$N=4$  $SU(n)$  Yang-Mills theory 
with the gauge group broken down to $U(1)^{n-1}$.
Our method of computing ${\cal H}(\bar{W},W)$
is equally powerful for arbitrary semi-simple gauge groups
and naturally leads to a nice algebrac structure
encoded in ${\cal H}(\bar{W},W)$.

It is interesting to note that ${\cal H}(\bar{W},W)$ is in general
unambiguously defined when $W$ lies along the flat directions of the $N=2$
Yang-Mills potential
\be
[{\bar W}, W] ~= ~0\;.
\label{flat}
\ee
Otherwise, the following identity \cite{3}
\be
\{ \cD^i_\a , \cD^j_\b \} W ~ =  ~ 2{\rm i}\,  \ve_{\a \b} \,\ve^{ij}\, [{\bar W}, W] 
\ee
implies that some higher derivative terms, which are denoted by the dots in  (\ref{1}), 
can also contribute to 
${\cal H}(\bar{W},W)$. Such problems do not appear when eq. (\ref{flat}) takes place.

As is well known, the most powerful approach to investigate quantum supersymmetric
field theories is to make use of an unconstrained superfield formulation.
Unfortunately, such a manifestly supersymmetric formulation
for the $N=4$ Yang-Mills theory is not known. 
For our present purpose, however, 
it is sufficient to realize the  $N=4$ Yang-Mills theory
as a theory of $N=2$ unconstrained superfields. 
The $N=2$ harmonic superspace \cite{7} is the only manifestly supersymmetric 
formalism developed to describe general $N=2$ Yang-Mills theories 
in terms of unconstrained (analytic) superfields.
This approach has been successfully applied for investigating effective action in
various $N=2$ supersymmetric models in recent papers
\cite{8,9,10,11,6}.

${}$From the point of view of $N=2$ supersymmetry, 
the $N=4$ Yang-Mills theory describes coupling of the $N=2$ vector multiplet
to the hypermultiplet in the adjoint representation. In the harmonic superspace
approach, the vector multiplet is realized by an unconstrained analytic gauge
superfield $V^{++}$. As concerns the hypermultiplet, it can be described either
by a real unconstrained analytic superfield $\o$  ($\o$-hypermultiplet) 
or by a complex unconstrained analytic superfield $q^+$ and its conjugate
$\breve{q}^+$ ($q$-hypermultiplet). In the $\o$-hypermultiplet realization, 
the classical action of $N=4$ Yang-Mills theory reads
\begin{equation}
S[V^{++},\omega]=\frac{1}{2g^2} {\rm tr} \int {\rm d}^4x {\rm d}^4\theta \,W^2-
\frac{1}{2g^2} {\rm tr} \int d\zeta^{(-4)}\,\nabla^{++}\omega
\nabla^{++}\omega
\label{3}
\end{equation}
where the second term describes the $\omega$-hypermultiplet action and
is integrated   over the analytic subspace of harmonic 
superspace (see Refs. \cite{7,10,6} for more details and notation). The
first term in (\ref{3}) is the pure $N=2$ Yang-Mills action. 
The explicit expression for 
the strength $W$ via
the prepotential $V^{++}$ 
is given in \cite{12}. 
The theory with action
(\ref{3}) is manifestly $N=2$ supersymmetric.  
However, the action (\ref{3}) turns out to be invariant
under two hidden supersymmetric transformations \cite{7}
\bea
\delta V^{++} &=& u^+_i \left( \e^{\a i} \q^+_\a  + {\bar \e}^i_\ad {\bar \q}^{+\ad} \right) \o \non \\
\d \o &=& 
-\frac{1}{8} u^-_i \left\{ (D^+)^2 (\e^i \q^- W_\l) 
+ ({\bar D}^+)^2 ({\bar \e}^i {\bar  \q}^- {\bar W}_\l )\right\}.
\eea
Here $W_\l$ denotes the strength in the $\l$-frame \cite{7,10}.
In the $q$-hypermultiplet reaization, the $N=4$ Yang-Mills theory is given by the action
\be
 S[V^{++},  q^+, \breve{q}^+]=\frac{1}{2g^2} {\rm tr} \int {\rm d}^4x {\rm d}^4\theta \,W^2-
\frac{1}{2g^2} {\rm tr} \int d\zeta^{(-4)}\, q^{+i} \nabla^{++} q^+_i 
\label{qaction}
\ee
where
\be
q^+_i = (q^+, \breve{q}^+)\;, \qquad q^{+i} = \ve^{ij} q^+_j = (\breve{q}^+, - q^+)\;.
\label{qhyp}
\ee
This model is manifestly $N=2$ supersymmetric. It also possesses two hidden supersymmetries
\bea
\delta V^{++} &=&  \left( \e^{\a i} \q^+_\a  + {\bar \e}^i_\ad {\bar \q}^{+\ad} \right) q^+_i \non \\
\d q^{+i} &=&  
-\frac{1}{4}  \left\{ (D^+)^2 (\e^i \q^- W_\l) 
+ ({\bar D}^+)^2 ({\bar \e}^i {\bar  \q}^- {\bar W}_\l )\right\}\;.
\eea

To provide manifest gauge invariance and supersymmetry at the quantum level,
we study the effective action for the classically equivalent theories (\ref{3}) and (\ref{qaction})
within the $N=2$ superfield
background field method \cite{10,11}. 
In accordance with \cite{10,6,8}, the one-loop effective action in both realizations is given by
\begin{equation}
\Gamma^{(1)}[V^{++}]=\frac{{\rm i}}{2}\,{\rm Tr}_{(2,2)}{\rm
ln}\stackrel{\frown}{\Box}-\frac{{\rm i}}{2}\,{\rm Tr}_{(4,0)}{\rm
ln}\stackrel{\frown}{\Box}
\label{4}
\end{equation}
where $\stackrel{\frown}{\Box}$ is the analytic d'Alambertian
introduced in \cite{10}
\bea
\stackrel{\frown}{\Box}&=&{\cal D}^m{\cal D}_m+\frac{{\rm i}}{2}({\cal D}^
{+\alpha}W){\cal D}^-_\alpha+
\frac{{\rm i}}{2}(\bar{\cal D}^+_{\dot{\alpha}}\bar{W})\bar{\cal
D}^{-\dot{\alpha}}-\frac{{\rm i}}{4}({\cal D}^{+\alpha}{\cal D}^+_\alpha W)
{\cal D}^{--} \non \\
&&+\frac{{\rm i}}{8}[{\cal D}^{+\alpha},{\cal D}^-_\alpha]W+
\frac{1}{2}\{\bar{W},W\}\;.
\label{5}
\eea
The formal definitions of the ${\rm Tr}_{(2,2)}{\rm
ln}\stackrel{\frown}{\Box}$
and ${\rm Tr}_{(4,0)}{\rm
ln}\stackrel{\frown}{\Box}$
are given in Ref. \cite{6}.

${}$For computing ${\cal H}(\bar{W},W)$ it is sufficient in fact to consider a special background
\begin{equation}
{\cal D}^{\alpha(i}{\cal D}_\alpha^{j)}W=0\;.
\label{6}
\end{equation}
Then, one can get  the following path integral
representation for $\Gamma^{(1)}$ \cite{6}
\begin{equation}
\exp \left({\rm i}\Gamma^{(1)}\right)=\frac{\int{\cal D}{\cal F}^{++}\exp\left\{
-\frac{{\rm i}}{2}{\rm tr}
\int {\rm d} \zeta^{(-4)}{\cal F}^{++}\stackrel{\frown}
{\Box}{\cal F}^{++}\right\}}
{\int{\cal D}{\cal F}^{++}\exp\left\{
-\frac{{\rm i}}{2}{\rm tr}\int {\rm d}\zeta^{(-4)}{\cal F}^{++}
{\cal F}^{++}\right\}}~.
\label{7}
\end{equation}
The superfield ${\cal F}^{++}(z,u)$ belonging to the adjoint representation 
looks like
${\cal F}^{++}(z,u)={\cal F}^{ij}(z)u^+_iu^+_j$, 
with ${\cal F}^{ij} ={\cal F}^{ji} $ satisfying
the constraints
\begin{equation}
{\cal D}^{(i}_\alpha{\cal F}^{jk)}=
\bar{\cal D}^{(i}_{\dot{\alpha}}{\cal F}^{jk)}=0\;, \qquad \overline{\cF^{ij}} = \cF_{ij}\;.
\label{8}
\end{equation}
The operator $\stackrel{\frown}{\Box}$ acts on ${\cal F}^{ij}$ as follows
\begin{equation}
\stackrel{\frown}{\Box}{\cal F}^{ij}=({\cal D}^m{\cal D}_m+\frac{1}{2}
\{\bar{W},W\}){\cal F}^{ij}+\frac{{\rm i}}{3}{\cal D}^{\alpha(i}W
{\cal D}_{\alpha |k|}{\cal F}^{j) k}+\frac{{\rm i}}{3}\bar{\cal D}^{(i}_
{\dot{\alpha}}\bar{W}\bar{\cal D}^{\dot{\alpha}}_{|k|}{\cal F}^{j)k}\;.
\label{9}
\end{equation}

Representation (\ref{7}) involves path integrals over constrained $N=2$ superfields.
Our aim now is to transform these path integrals to those over unconstrained 
$N=1$ superfields. We introduce $N=1$
Grassmann coordinates $(\theta^\alpha,\bar{\theta}_{\dot{\alpha}})$
by the rule $\theta^\alpha=\theta^\alpha_1$,
$\bar{\theta}_{\dot{\alpha}}=\bar{\theta}^1_{\dot{\alpha}}$, 
the corresponding gauge covariant derivatives
${\cal D}_\alpha={\cal D}_\a^1$,
$\bar{\cal D}^{\dot{\alpha}}=\bar{\cal D}^{\dot{\alpha}}_1$
and then define
the $N=1$ projection of an arbitrary $N=2$ superfield $f(z^M)$ by the standard rule
$f|=f(x^m,\theta^\alpha_i,\bar{\theta}^i_{\dot{\alpha}})|_{\theta_2=
\bar{\theta}^2=0}$.  As is well known, from the $N=2$ Yang-Mills strength $W$ 
one obtains two $N=1$
covariantly chiral superfields $\F = W|$ and  $2{\rm i} W_\a = {\cal D}^2_\a W|$.
The $N=1$ projections of ${\cal F}^{ij}$ read
\begin{equation}
\Psi={\cal F}^{22}|\,,\qquad \bar{\Psi}={\cal F}^{11}|\,,\qquad
F=\bar{F}=-2i{\cal F}^{12}|
\label{10}
\end{equation}
and satisfy the constraints
\begin{equation}
\bar{\cD}_{\dot{\alpha}}\Psi=0\,,\qquad -\frac{1}{4}\bar{\cal
D}^2F+[\Phi,\Psi]=0
\label{11}
\end{equation}
Therefore, $\Psi$ is a covariantly chiral $N=1$ superfield while the real superfield
$F$ is subject to a modified linear constraint.

Until this point, the $N=2$ Yang-Mills strength was constrained only by eq. (\ref{6}).
Now, we specify $W$ to belong to the Cartan
subalgebra and, hence, to satisfy eq. (\ref{flat}). 
Moreover, we require the $N=1$ components 
of $W$ to be (covariantly) constant,
\be
{\cal D}_\alpha\Phi=0\,,\qquad {\cal D}_\alpha W_\beta=0\;.
\ee
Such a  background is still sufficient for calculating 
${\cal H}(\bar{W},W)$, since the identity
\be
\int {\rm d}^{4}x  {\rm d}^8 \q \; \cH(\bar W, W) ~=~ 
\int {\rm d}^{8} z\; W^\a W_\a {\bar W}_\ad {\bar W}^\ad \;
\frac{\pa^4 \cH(\bar \F, \F)}{\pa \F^2 \pa {\bar \F}^2} ~~+ ~~
{\rm derivatives}
\label{derivatives}
\ee
along with the requirement of scale and chiral invariance
allow us to uniquely restore ${\cal H } (\bar{W},W)$ .
Here ${\rm d}^8 z$ denotes the full $N=1$ superspace measure.
For the background chosen
the operator $\stackrel{\frown}{\Box}$ does not mix the superfields
$\Psi$, $\bar{\Psi}$ and $F$ 
\begin{equation}
\Delta ({\cal F}^{ij}|)=(\stackrel{\frown}{\Box}{\cal F}^{ij})|
\label{12}
\end{equation}
where
\begin{equation}
\Delta={\cal D}^m{\cal D}_m-W^\alpha{\cal D}_\alpha+\bar{W}_{\dot
{\alpha}}\bar{\cal D}^{\dot{\alpha}}+\frac{1}{2}\{\Phi,\bar{\Phi}\}\;.
\label{13}
\end{equation}
Expressing the $N=2$ integration variables in  (\ref{7}) via
their $N=1$  projections, we obtain the following
representation for  $\Gamma^{(1)}$ in terms of path
integrals over (still constrained) $N=1$ superfields
\footnote{It is worth  pointing out that we deduce
eq. (\ref{14}) from the representation  (\ref{7})
for $\G^{(1)}$ which is manifestly $N=2$ supersymmetric 
and invariant with respect to the automorphism 
$SU(2)_R$ symmetry. That is why it is in our power 
to make use of any useful technique in order to 
compute special contributions to  $\G^{(1)}$,
in particular, to reduce  $\G^{(1)}$ to $N=1$
superfields. This is completely different to the 
case when the $N=2$ or $N=4$ theories are 
formulated from the very beginning in $N=1$
superfields, when only $N=1$ supersymmetry
is realized off-shell; in such a case the effective
action possesses  $N=1$ supersymmetry only.
By construction, our approach is manifestly 
$N=2$ supersymmetric,
in spite of the comments given in \cite{14}.}
  \begin{equation}
\exp \left( {\rm i} \Gamma^{(1)} \right)
~=~\frac{\int{\cal D}\bar{\Psi}{\cal D}\Psi \cD F
\exp\left\{ {\rm i} ~{\rm tr}
\int {\rm d}^8z (-\bar{\Psi}\Delta\Psi+\frac{1}{2}
F\Delta F)\right\}}
{\int{\cal D}\bar{\Psi}{\cal D}\Psi \cD F
\exp\left\{ {\rm i} ~{\rm tr}
\int {\rm d}^8 z (-\bar{\Psi}\Psi+\frac{1}{2}F^2)\right\}}\;.
\label{14}
\end{equation}
Our next  step is to evaluate the right hand side of
eq. (\ref{14}).

Until now the gauge group was completely arbitrary. Let  us 
specialize our consideration to the case  of $SU(n)$. To start with
we make a quick tour through the corresponding Lie algebra $su(n)$ 
\cite{13}  consisting of hermitian traceless matrices. We
introduce the Weyl basis $\{e_{kl}\}$ of $su(n)$ \footnote{From now on,
small Latin letters are used for $SU(n)$ indices.}
\begin{equation}
(e_{kl})_{pq}=\delta_{kp}\delta_{lq}\,, \qquad \quad k,l,p,q=1,2,\dots,n
\label{15}
\end{equation}
Then an  arbitrary element $a \in su(n)$ looks like
\begin{equation}
a=\sum\limits_{k=1}^na^ke_{kk}+\sum_{k \neq l}   a^{kl}e_{kl}\,,
\qquad a^{kl}=\overline{a^{lk}} \,, \qquad
\sum\limits_{k=1}^n a^k=0
\label{16}
\end{equation}
with  $a^i$ being real. The elements $r$ of the Cartan subalgebra are
\begin{equation}
r=\sum\limits_{k=1}^n r^ke_{kk}=\mbox{diag}(r^1,r^2,\dots,r^n)\,,
\qquad \sum\limits_{i=1}^n r^i=0\;.
\label{17}
\end{equation}
${}$For any elements of the Weyl basis we have
\begin{equation}
{\rm tr}(e_{pq}e_{kl})=2n{\rm tr}_F(e_{pq}e_{kl})=2n\,\delta_{pl}\,
\delta_{qk}\;.
\label{18}
\end{equation}
Here `${\rm tr}_F$' denotes the trace in the fundamental representation. 
From here one gets important consequences
\begin{equation}
{\rm tr}(e_{kl}e_{lk})=2n\;;\qquad \qquad
{\rm tr}(e_{pq}e_{kl})=0\,,\qquad p\neq l,\ q\neq k\;.
\label{19}
\end{equation}
Given an element $r$ of the Cartan subalgebra, one finds
\begin{equation}
[r,e_{kl}]=(r^k-r^l)e_{kl}
\label{20}
\end{equation}
with  the eigenvalues $(r^k-r^l)$  defining the roots of $su(n)$.

${}$For the gauge group chosen, the strengths $W$ and $\bar{W}$
lie in the Cartan subalgebra of $su(n)$ 
\be
W={\rm diag}(W^1,W^2,\dots,W^n) \;, \qquad
\sum\limits_{k=1}^nW^k=0\;.
\label{strength}
\ee 
Since we are interested in the situation  when the gauge group $SU(n)$ 
is  broken down to the maximal torus 
$U(1)^{n-1}$, we should have
$W^k-W^l\neq 0$ for $k \neq l$. 
In the opposite case, when  several eigenvalues $W^k$ coincide,
some nonabelian group $H \in SU(n)$ remains unbroken. 
Introducing the $N=1$ projections $\Phi=W|$ and
$W_\alpha=-\frac{{\rm i}}{2}{\cal D}^2_\alpha W|$ associated with  $W$, 
we obtain the $N=1$
superfield roots $\Phi^k-\Phi^l$ and $W^k_\alpha-W^l_\alpha$.
The above restrictions on $W^{k}$ are equivalent to 
$\Phi^k-\Phi^l\neq 0$ for $k \neq l$. 

Let  us return to eq. (\ref{14}). Since the strengths $\F$ and  $W_\a$
belong to the Cartan subalgebra,
the components of the quantum superfields
$\bar{\Psi}$, $\Psi$, $F$ which lie in the Cartan subalgebra do not interact
with the background field and therefore they {\it completely decouple}. 
On the other hand, 
the components of $\Psi$ and $\bar{\Psi}$ out of the Cartan
subalgebra {\it are expressed} via $F$ and $\bar F$  with the aid of 
constraints (\ref{11}) 
\begin{equation}
\Psi^{kl}=\frac{\bar{\cal D}^2F^{kl}}{4(\Phi^k-\Phi^l)}\,,\qquad
\bar{\Psi}^{kl}=\frac{{\cal D}^2F^{kl}}{4(\bar{\Phi}^k-\bar{\Phi}^l)}\;,
\label{21}
\end{equation}
and these expressions are nonsingular in the case under
consideration. As a result, we can transform the right hand side
of eq.  (\ref{14}) to path integrals over {\it unconstrained superfields}
\be
V^{kl} \equiv F^{kl} \;, \qquad {\bar V}^{kl} \equiv F^{lk}\;, \qquad \qquad 
k < l \;.
\ee

Taking into account eqs. (\ref{19}) and (\ref{21}), 
we can transform the integral in the
denominator of eq. (\ref{14}) as follows
\begin{equation}
{\rm tr} \int {\rm d}^8z(-\bar{\Psi}\Psi+\frac{1}{2}F^2)=
2n\int {\rm d}^8 z \sum_{k<l} \bar{V}^{kl}B_{kl}V^{kl}
\label{22}
\end{equation}
where
\begin{equation}
B_{kl}=\frac{1}{16}\frac{\{\bar{\cal D}^2,{\cal D}^2\}}{|\Phi^k-
\Phi^l|^2}+1\:.
\label{23}
\end{equation}
It is worth pointing out that the sum in
(\ref{22}) is taken over half the roots and we can choose 
the positive roots to contribute to (\ref{22}). As a result
\begin{eqnarray}
\int {\cal D}\bar{\Psi}{\cal D}\Psi{\cal D}F & \exp & \left\{{\rm i}\,
{\rm tr}
\int {\rm d}^8z(-\bar{\Psi}\Psi+\frac{1}{2}F^2)\right\}\nonumber\\
=\int{\cal D}\bar{V}^{kl}{\cal D}V^{kl} & \exp & \Big\{2n \,{\rm i}
\int {\rm d}^8z\sum\limits_{k<l}\bar{V}^{kl}B_{kl}V^{kl}\Big\}
=\prod\limits_{k<l}{\rm Det}^{-1}(B_{kl}) \;.
\label{24}
\end{eqnarray}
Next we  turn to the nominator in  (\ref{14}). First of all we
find the action of $\Delta$ (\ref{13}) on the superfields
$F^{kl}$. The result reads
\begin{equation}
\Delta(F^{kl}e_{kl})=(\Delta_{kl}F^{kl})e_{kl} \qquad \quad ({\rm no ~~sum})
\label{25}
\end{equation}
where
\begin{equation}
\Delta_{kl}={\cal D}^m{\cal D}_m-(W^{k\alpha}-W^{l\alpha}){\cal
D}_\alpha+(\bar{W}^k_{\dot{\alpha}}-\bar{W}^l_{\dot{\alpha}})
\bar{\cal D}^{\dot{\alpha}} +|\Phi^k-\Phi^l|^2\;.
\label{26}
\end{equation}
Using (\ref{21}) and fulfilling  straightforward calculations we get
\begin{equation}
{\rm tr} \int {\rm d}^8z\left( \hf F\Delta F-\bar{\Psi}\Delta
\Psi \right)=2n  \int {\rm d}^8z \sum\limits_{k<l}{\bar V}^{kl}
B_{kl}\Delta_{kl}V^{kl}
\label{27}
\end{equation}
where $B_{kl}$ is given by eq. (\ref{23}).
From here we   obtain
\begin{eqnarray}
\int{\cal D}\bar{\Psi}{\cal D}\Psi{\cal D}F & \exp & \left\{ {\rm i} 
\,{\rm tr}
\int {\rm d}^8z \left(-\bar{\Psi}\Delta\Psi+\frac{1}{2}F\Delta F\right)
\right\}\nonumber\\
=\int{\cal D}\bar{V}^{kl}{\cal}V^{kl} & \exp & \Big\{ 2 n \,{\rm i} 
\int {\rm d}^8z\sum\limits_{k<l}\bar{V}^{kl}B_{kl}\Delta_{kl}V^{kl}\Big\}
=\prod\limits_{k<l}{\rm Det}^{-1}(B_{kl}){\rm Det}^{-1}(\Delta_{kl})\;.
\label{28}
\end{eqnarray}
We result with 
\begin{equation}
{\rm e}^{{\rm i} \Gamma^{(1)}}=\prod\limits_{k<l}{\rm Det}^{-1}(\Delta_{kl})\;.
\label{29}
\end{equation}
It is seen that the one-loop correction $\Gamma^{(1)}$ to effective action
is determined by the functional determinant of the operator (\ref{26}) on
the space of unconstrained $N=1$ superfields under the Feynman boundary 
conditions. Eq. (\ref{29}) can be rewritten as follows
\begin{equation}
\Gamma^{(1)}=\sum\limits_{k<l}\Gamma_{kl}\;,\qquad
\Gamma_{kl}={\rm i}\, {\rm Tr} \; {\rm ln}\Delta_{kl}\;.
\label{30}
\end{equation}

The $SU(n)$-operator $\Delta_{kl}$ (\ref{26}) has the same structure as 
the $SU(2)$-operator $\Delta$ introduced  in \cite{6}. Therefore we
can apply the technique developed in  \cite{6} and obtain
\begin{equation}
\Gamma_{kl}=\frac{1}{(4\pi)^2}\int {\rm d}^8z\frac{W^{\alpha kl}W^{kl}_\alpha
\bar{W}^{kl}_{\dot{\alpha}}\bar{W}^{\dot{\alpha}kl}}
{(\Phi^{kl})^2(\bar{\Phi}^{kl})^2}
\label{31}
\end{equation}
where
\begin{equation}
\Phi^{kl}=\Phi^k-\Phi^l\;, \qquad
W^{kl}_\alpha=W^k_\alpha-W^l_\alpha \;.
\label{32}
\end{equation}
Eqs. (\ref{30}--\ref{32}) define the non-holomorphic effective potential
${\cal H}(\bar{W},W)$ of the $N=4$ Yang-Mills theory in terms of 
the $N=1$ projections of $W$ and $\bar{W}$. 

${}$From eqs. (\ref{derivatives}) and (\ref{30}--\ref{32})  one can easily restore 
${\cal H}(\bar{W},W)$:
\bea
\Gamma^{(1)}& =& \int {\rm d}^{4}x {\rm d}^8 \q \;{\cal H}(\bar{W},W) \non \\
{\cal H}(\bar{W},W) & = & 
\frac{1}{(8\pi)^2} \sum_{k<l} 
\ln \left( \frac{ {\bar W}^{k}  - {\bar W}^{l}  }{\L} \right) ^2
\ln \left( \frac{W^{k}  - W^l}{\L} \right)^2
\label{final}
\eea
where the strengths $W^k$ are chosen as in  (\ref{strength}), with $W^k -W^l \neq 0$
for $k \neq l$. 
Eq. (\ref{final}) is our final result.
Similar to the holomorphic effective potential $\cF(W)$ \cite{2},
the non-holomorphic effective potential  is
constructed in terms of the roots of $SU(n)$ and  obviously
invariant under the Weyl group. 
Some bosonic contributions to $\G^{(1)}$ were discussed 
in \cite{tseytlin}.

It is necessary to point out that our method to compute
the non-holomorphic effective
potential ${\cal H}(\bar{W},W)$ is general and perfectly works
for arbitrary semi-simple gauge groups, for instance, $SO(n)$.
The starting point is  representation  (\ref{7}).
Then, one has to specify the Cartan subalgebra and Weyl basis
for the gauge group in field 
and, finally, it remains to repeat the technical steps described.
Given a semi-simple rank-$r$ gauge group $G$, we introduce
its Weyl basis $\{ h_{\hat{i}} ,~ e_{+ \hat{ \a}},~ e_{- \hat{\a}} \}$,
where the elements $h_{\hat{i}}$ span the Cartan subalgebra, 
$\hat{i} = 1,\dots, r$, and $\pm \hat{\a}$ are the positive (negative) roots.
When the gauge group is broken down to its maximal torus $U(1)^r$,
the $N=2$ strength looks like $W= \sum W_{\hat{i}} h_{\hat{i}}$,
$[W, e_{+ \hat{\a}}] = W_{+ \hat{ \a}}\, e_{+ \hat{ \a}}$, with
all $W_{+ \hat{ \a}}$ being non-vanishing. The non-holomorphic 
effective potential reads
\bea
{\cal H}(\bar{W},W)  &=&  
\frac{1}{(8\pi)^2} 
\sum\limits_{ 
{\rm pos. ~roots}
} 
\ln \Big( \frac{ {\bar W}_{+\hat{\a}}  }{\L} \Big) ^2\;
\ln \Big( \frac{W_{+\hat{\a}} }{\L} \Big)^2
\eea
and this is similar to the structure
of perturbative holomorphic effective potential \cite{2}.

When this work was completed, 
there appeared recent papers \cite{14,15} where similar
results were obtained by different methods.
\vspace{5mm}

\noindent
{\bf Acknowledgements.}
The authors are grateful to E.A. Ivanov, B.A. Ovrut and S. Theisen  for
valuable discussions. We are grateful to A.A. Tseytlin for bringing 
Ref. \cite{tseytlin} to our attention.
We acknowledge a partial support from INTAS
grant, INTAS-96-0308. I.L.B. and S.M.K. are grateful to RFBR grant,
project No 96-02-16017 and RFBR-DFG grant,
project No 96-02-00180 for partial support.

\end{document}